\documentclass[a4paper,oneside,final,notitlepage,onecolumn,12pt]{article}
\usepackage{amsfonts}
\usepackage{epsf}
\usepackage{graphicx}% Include figure files
\usepackage{amssymb,eso-pic}
\usepackage{latexsym}
\usepackage{tabularx}
\usepackage{amsxtra} 
\usepackage{hyperref}
\usepackage{t1enc}
\usepackage{amsmath,accents}

\usepackage{cancel}

\usepackage{ifpdf,epsfig,array,amsmath,amssymb}

\usepackage{psfrag} 
\psfrag{na}{\footnotesize $n^a$}   
\psfrag{ta}{\footnotesize $\sigma^a$}   
\psfrag{vna}{\footnotesize $\check{n}^a$}
\psfrag{vNa}{\footnotesize $\check{N}^a$}
\psfrag{scrW}{\footnotesize $\mycal{W}_{\rho_0}$}
\psfrag{t=t1}{\footnotesize $\Sigma_{\sigma_1}$}
\psfrag{t=t2}{\footnotesize $\Sigma_{\sigma_2}$}
\psfrag{r=all}{\footnotesize $\rho=const$}
\psfrag{hna}{\footnotesize $\hat{n}^a$}
\psfrag{tna}{\footnotesize $\tilde{n}^a$}
\psfrag{hab,Kab}{\footnotesize $h_{ab}, K_{ab}$}
\psfrag{thab,tKab}{\footnotesize $\tilde h_{ab}, \tilde K_{ab}$}
\psfrag{hhab,hKab}{\footnotesize $\hat h_{ab}, \hat K_{ab}$}
\psfrag{chab,cKab}{\footnotesize $\check{h}_{ab}, \check{K}_{ab}$}

\usepackage[normalem]{ulem}
\usepackage{color}
\definecolor{blue}{rgb}{0,0,1}
\definecolor{red}{rgb}{1,0,0}

\definecolor{DGREEN}{rgb}{0,0.7,0.3}
\definecolor{grey1}{rgb}{0.52, 0.52, 0.51}

\newcommand{\interior}[1]{\accentset{\smash{\raisebox{-0.12ex}{$\scriptstyle\circ$}}}{#1}\rule{0pt}{2.3ex}}
\makeatother

\makeatletter
\def\@xfootnote[#1]{%
  \protected@xdef\@thefnmark{#1}%
  \@footnotemark\@footnotetext}
\makeatother

\DeclareFontFamily{OT1}{rsfs}{} \DeclareFontShape{OT1}{rsfs}{m}{n}{
<-7> rsfs5 <7-10> rsfs7 <10-> rsfs10}{}
\DeclareMathAlphabet{\mycal}{OT1}{rsfs}{m}{n}

\def\sc{{\hskip 3.5pt {{}^{{}^{{}_{{}_{\bowtie}}}}} \kern -8.pt{}}}  
\def\SC{{\hskip 3.5pt {{}^{{}^{{}^{{}_{{}_{\bowtie}}}}}} \kern -10.5pt{}}}

\DeclareMathAlphabet{\mathpzc}{OT1}{pzc}{m}{it}

\newcommand{\hoch}[1]{$\, ^{#1}$}

\newcommand{\auth}{{Istv\'an R\'{a}cz\hoch{\S,}\,\footnote[$\sharp$]{~email: racz.istvan@wigner.mta.hu}
 \ and \ Jeffrey Winicour\hoch{\flat,}\,\footnote[$\natural$]{~email: winicour@pitt.edu} }}

\begin{document}

%%%%%%%%%%%%%%%%%%%%%%%%%%%%%%%%%%%%%%%%%%%%%%
\newtheorem{theorem}{Theorem}[section]
\newtheorem{lemma}{Lemma}[section]
\newtheorem{proposition}{Proposition}[section]
\newtheorem{corollary}{Corollary}[section]
\newtheorem{conjecture}{Conjecture}[section]
\newtheorem{example}{Example}[section]
\newtheorem{definition}{Definition}[section]
\newtheorem{remark}{Remark}[section]
\newtheorem{exercise}{Exercise}[section]
\newtheorem{axiom}{Axiom}[section]
%%%%%%%%%%%%%%%%%%%%%%%%%%%%%%%%%%%%%%%%%%%%%
\renewcommand{\theequation}{\thesection.\arabic{equation}}

\begin{center}

{\LARGE{\bf Black hole initial data without elliptic equations}}

\vspace{25pt}
\auth

\vspace{30pt}{\hoch{\S}\it Wigner RCP, \\ 
H-1121 Budapest, Konkoly Thege Mikl\'os \'ut 29-33., Hungary}

\vspace{10pt}{\hoch{\flat}\it Department of Physics and Astronomy,\\ 
University of Pittsburg, Pittsburgh, PA, 15260,  USA}

\begin{abstract}

We explore whether a new method to solve the constraints of Einstein's equations,
which does not involve elliptic equations,
can be applied to provide initial data for black holes.  We show that this method
can be successfully applied to a nonlinear
perturbation of a Schwarzschild black hole by establishing
the well-posedness of the resulting constraint problem.
We discuss its possible generalization
to the boosted, spinning multiple black hole problem.
  
\end{abstract} 

\end{center}

%%%%%%%%%%%%%%%%%%%% INTRODUCTION %%%%%%%%%%%

\section{Introduction}
\label{introduction}

The prescription of physically realistic initial data for black holes is a crucial ingredient to the simulation
of the inspiral and merger of binary black holes and the computation or the radiated gravitational waveform.
Initialization of the simulation is a challenging
problem due to the nonlinear constraint equations that the data must satisfy. The traditional solution
expresses the constraints in the form of elliptic equations. Here we consider a radically new method of
solving the constraints which does not require elliptic solvers~\cite{racz_constraints}.
We show, at least for
nonlinear perturbations of Schwarzschild black hole data, that the Hamiltonian and momentum constraints 
lead to a well-posed
strongly hyperbolic problem whose solutions
satisfy the full constraint system.
The possibility of extending this approach to binary black holes offers a simple alternative
way to provide boundary
conditions for the initialization problem that might prove to be more physically realistic.

The inspiral and merger of  a binary black hole
is expected to be the strongest possible source of gravitational radiation
for the emerging field of gravitational wave astronomy. The details of the gravitational
waveform supplied by numerical simulation is a key tool to enhance detection of the gravitational
signal and interpret its scientific content. It is thus important that the initial data does
not introduce spurious effects into the waveform. Such ``junk radiation'' is common to
all current methods for supplying initial data and appears early in the simulation
as a high frequency component of the waveform.
This can  be a troublesome feature with regard to matching
the waveform in the nonlinear regime spanned by the simulation to the post-Newtonian
chirp waveform provided by perturbation theory. The initial parameters governing
the black hole spins, mass ratio and ellipticity of the binary orbit have to be adjusted to include the effect of
this transitory period. As a result, it becomes difficult to match exactly to the parameters
governing the post-Newtonian orbit.
In addition, although the high frequency component of the junk radiation appears to
dissipate after
some early transitory period,
there is no quantitative measure of its low frequency component which might
affect the ensuing waveform.

All initialization methods presently in use reduce the constraint problem
to a system of elliptic equations, which require boundary conditions at inner boundaries
in the strong field region
surrounding the singularities inside the black holes, as well as at an outer boundary
surrounding the system. The new method we consider here only requires
data on the outer boundary, which is in the weak field region where the choice of
boundary data can be guided by asymptotic flatness.
The constraints are then satisfied by an  inward ``evolution''
of the hyperbolic system
along radial streamlines.

The initial data for solving Einstein's equations consist of a pair of symmetric tensor fields $(h_{ij},K_{ij})$
on a smooth three-dimensional manifold $\Sigma$, where $h_{ij}$ is a Riemannian metric and
$K_{ij}$ is interpreted as the extrinsic curvature of $\Sigma$  after its embedding
in a 4-dimensional space-time.
The constraints on a vacuum solution (see e.g.~Refs.~\cite{choquet,wald})
consist of
\begin{align} 
{}^{{}^{(3)}}\hskip-1mm R + \left({K^{j}}_{j}\right)^2 - K_{ij} K^{ij} =0\,, \label{expl_eh}\\
D_j {K^{j}}_{i} - D_i {K^{j}}_{j} =0\,,\label{expl_em}
\end{align}
where ${}^{{}^{(3)}}\hskip-1mm R$ and $D_i$ denote the scalar curvature and the
covariant derivative operator associated with $h_{ij}$, respectively. 

The standard approach to solving the constraints is based upon the conformal method,
introduced by Lichnerowicz~\cite{lich} to recast the Hamiltonian constraint (\ref{expl_eh})
as an elliptic equation and later extended by York~\cite{york0,york1} to reduce the momentum
constraint (\ref{expl_em}) also to an elliptic system. For a review of the historic implementation
of this method in numerical relativity see~\cite{cook}.

A major obstacle in prescribing black hole initial data is the presence of a singularity
inside the black hole. The initial strategy for handling the singularity was the excision of
the singular region inside the black hole~\cite{excision}. In this case, an artificial
inner boundary condition for the elliptic system is posed on boundaries inside the
apparent horizons surrounding  the individual black holes.
Other strategies have since been proposed.
One is the puncture method in which the initial hypersurface extends though a wormhole
to an internal asymptotically flat spatial infinity, which is then
treated by conformal compactification~\cite{puncture}. Here the
freedom in the choice of conformal factor governing the compactification
enters as an effective boundary condition. In addition, it is known that
the puncture quickly changes its nature. In fact, early attempts to simulate binary
black holes failed until it was realized that the punctures must be allowed to move.
Studies of this feature in the case of a single black hole revealed
that the puncture quickly transits from the internal spatial infinity to an internal
timelke infinity~\cite{trumpet1}. This realization has given rise to the
trumpet version of initial data,  in which the initial
Cauchy hypersurface extends to an internal timelike infinity with asymptotically
finite surface area~\cite{trumpet1,trumpet2}. Trumpet data offers a
promising alternative to puncture data but its merits have not yet been extensively explored
in binary black hole simulations~\cite{hannam}.

Coupled to these techniques for avoiding singularities is the choice of initial time slice.
For example, there are many ways to prescribe Schwarzschild initial data depending, say, 
upon whether the initial Cauchy hypersurface is time symmetric or
horizon penetrating. Here we will focus on initial data in Kerr-Schild form~\cite{ks1,ks2},
which for the Schwarzschild case corresponds to ingoing Eddington-Finklestein coordinates,
which extend from spatial infinity to the singularity
and penetrate the horizon. The new approach to solving the constraints that we consider
becomes degenerate for a time symmetric initial slice, whose extrinsic curvature vanishes.
However, time symmetric space times contain as much ingoing as outgoing
gravitational waves, so they are not the appropriate physical models for
studying binary waveforms. Although our focus here is on
data in Kerr-Schild form, we do not wish to imply that this approach
would not work for puncture or trumpet data.

A very attractive feature of Kerr-Schild initial data is that it provides a preferred Minkowski
background to construct boosted black holes by means of a Lorentz transformation.
Two independent ways of prescribing Kerr-Schild initial data have been
proposed. In one version, the 4-dimensional  aspect of the
 Kerr-Schild ansatz is preserved as
much as possible~\cite{ksb}.
This leads to a workable scheme for superimposing non-spinning black holes
but the generalization to the spinning case remains problematic. In the other case,
the Kerr-Schild ansatz is loosened to a 3-dimensional  version that allows superposition of
multiple spinning black holes~\cite{ksm1}. This has been implemented to
provide data for boosted, spinning binary black holes and plays
an important role in current simulations~\cite{ksm2}.

There are several variants to the new method of solving the constraints proposed
in \cite{racz_constraints, racz_geom_det, racz_geom_cauchy,racz_tdfd},
depending upon which components
of the initial data are assigned freely. They all avoid elliptic equations.
Here we apply the simplest of these variants to the initial data problem for black holes. In this variant,
the Hamiltonian and momentum constraints constitute a strongly hyperbolic system
which only requires data
on a 2-surface surrounding the black holes.

In Sec.~\ref{sec:setup}, we review this new approach.
In Sec.~\ref{sec:kerr-schild}, we show that the requirements for well-posedness
of the underlying algebraic-hyperbolic constraint problem are satisfied by a
Schwarzschild black hole described in Kerr-Schild form.
In Sec.~\ref{sec:pert}, we present an explicit proof that
nonlinear perturbations of Schwarzschild black hole data in Kerr-Schild form
lead to a well-posed strongly hyperbolic problem.

In Sec.~\ref{sec:conc}, we conclude with a discussion of the possibility of extending this
approach to general data for a system of boosted, spinning multiple black holes.
We show how the initial metric data for multiple black holes can
be freely prescribed in 4-dimensional superimposed  Kerr-Schild
form for the individual boosted, spinning black holes.
Two pieces of extrinsic curvature data, which represents the two gravitational degrees
of freedom, can also be freely prescribed by superimposing the individual black hole data.
The remaining extrinsic curvature data is then
determined by the algebraic-hyperbolic constraint system. In a linear theory, the
superposition of such non-radiative data would lead to a non-radiative solution.
This suggests that this new method may offer an alternative approach to suppressing
junk radiation and to controlling the effect of initial data on a binary orbit.
However, due to the nonlinearity of
Einstein's equations, there is no guarantee that, in the strong field region
between the individual black holes, this superimposed
free data does not introduce spurious radiation. A completely analytic resolution
of these issues does not seem possible.
A major motivation for this paper is to encourage the numerical experimentation
necessary to explore the merit and feasibility of this new approach.

\section{A new approach to the constraints}
\label{sec:setup}

\setcounter{equation}{0}

We assume that the topology of $\Sigma$ allows a  smooth foliation
by a one-parameter family of homologous two-surfaces.
In the application to black hole initial data,
we assume for simplicity a foliation $\mycal{S}_\rho$ by topological spheres 
described by the level surfaces $\rho=const$ of a smooth function.

Choose now a vector field $\rho^i$ on $\Sigma$ such that $\rho^i \partial_i \rho=1$.
Then the unit normal $\hat n^i$ to $\mycal{S}_\rho$ has the decomposition
\begin{equation}\label{nhat}
\hat n^i={\hat{N}}^{-1}\,[\, \rho^i-{\hat N}{}^i\,]\,,
\end{equation}
where the `lapse' $ \hat N$ and `shift' $\hat N^i$ of the vector field $\rho^i$
are determined by
$\hat n_i= \hat N \partial_i \rho$ and $\hat N^i=\hat \gamma{}^i{}_j\,\rho^j$,
with $\hat \gamma{}^i{}_j=\delta{}^i{}_j-\hat n{}^i\hat n_j$. 

The 3-metric $h_{ij}$ on $\Sigma$ then has the $2+1$ decomposition
\begin{equation}\label{hij}
       h_{ij}=\hat \gamma_{ij}+\hat  n_i \hat n_j\,,
\end{equation}
where $\hat \gamma_{ij}$ is the metric induced on the surfaces $\mycal{S}_\rho$.
The extrinsic curvature $\hat K_{ij}$ of $\mycal{S}_\rho$
is given by
\begin{equation}\label{hatextcurv}
     \hat K_{ij}= {{\hat \gamma}^{l}}{}_{i}\, D_l\,\hat n_
         j=\tfrac12\,\mycal{L}_{\hat n} {\hat \gamma}_{ij}\,.
\end{equation}

The extrinsic curvature $K_{ij}$ of $\Sigma$, which forms part of the initial data, 
has the decomposition
\begin{equation}
        K_{ij}= \boldsymbol\kappa \,\hat n_i \hat n_j  + \left[\hat n_i \,{\rm\bf k}{}_j  
        + \hat n_j\,{\rm\bf k}{}_i\right]  + {\rm\bf K}_{ij}\,,
\end{equation}
where $\boldsymbol\kappa= \hat n^k\hat  n^l\,K_{kl}$,
${\rm\bf k}{}_{i} = {\hat \gamma}^{k}{}_{i} \,\hat  n^l\, K_{kl}$
and ${\rm\bf K}_{ij} = {\hat \gamma}^{k}{}_{i} {\hat \gamma}^{l}{}_{j}\,K_{kl}$. 
Here we use boldfaced symbols to indicate tensor fields
tangent to $\mycal{S}_\rho$.
In addition, we shall denote the trace and trace free parts of $\hat K_{ij}$ and
${\rm\bf K}_{ij}$ by ${\hat K}{}^l{}_{l}=\hat\gamma^{kl}\,{\hat K}_{kl}$,
${\rm\bf K}^l{}_{l}=\hat\gamma^{kl}\,{\rm\bf K}_{kl}$,
$\interior{\hat K}{}_{ij}={\hat K}_{ij}-\tfrac12\,\hat \gamma_{ij}\,{\hat K}{}^l{}_{l}$
and  $\interior{\rm\bf K}_{ij}={\rm\bf K}_{ij}-\tfrac12\,\hat \gamma_{ij}\,{\rm\bf K}^l{}_{l}$,
respectively. 

By replacing the initial data set $(h_{ij},K_{ij})$ by the seven fields 
$(\hat N,\hat N^i,\hat \gamma_{ij}, \interior{\rm\bf K}_{ij}, \boldsymbol\kappa,{\rm\bf k}{}_{i}, {\rm\bf K}^l{}_{l})$,
the Hamiltonian and momentum constraints (\ref{expl_eh}) and (\ref{expl_em})
can be expressed as \cite{racz_constraints} 
(see also  \cite{racz_geom_det,racz_geom_cauchy,racz_tdfd})
\begin{align} 
                  \mycal{L}_{\hat n}({\rm\bf K}^l{}_{l}) - \hat D^l {\rm\bf k}_{l} 
                  + 2\,\dot{\hat n}{}^l\, {\rm\bf k}_{l}
                   - [\,\boldsymbol\kappa-\tfrac12\, ({\rm\bf K}^l{}_{l})\,]\,
                   ({\hat K^{l}}{}_{l})  + \interior{\rm\bf K}{}_{kl}\interior{\hat K}{}^{kl}  = {}& 0 \,,   
                   \label{constr_mom2}  \\
                     \mycal{L}_{\hat n} {\rm\bf k}{}_{i}  
              + ({\rm\bf K}^l{}_{l})^{-1}[\,\boldsymbol\kappa\,\hat D_i ({\rm\bf K}^l{}_{l})
               -2\, {\rm\bf k}{}^{l}\hat D_i{\rm\bf k}{}_{l}\,] 
              + (2\,{\rm\bf K}^l{}_{l})^{-1}\hat D_i\,[{}^{{}^{(3)}}\hskip-1mm R 
              - \interior{\rm\bf K}{}_{kl}\,\interior{\rm\bf K}{}^{kl} \,] {}& \nonumber \\ 
                 + ({\hat K^{l}}{}_{l})\,{\rm\bf k}{}_{i}
                  + [\,\boldsymbol\kappa-\tfrac12\, ({\rm\bf K}^l{}_{l})\,]\,\dot{\hat n}{}_i  
                 - \dot{\hat n}{}^l\,\interior{\rm\bf K}_{li}  
                 + \hat D^l \interior{\rm\bf K}{}_{li}  = {}& 0 \, , \label{constr_mom1}
\end{align}
where $\boldsymbol\kappa$ is determined by
\begin{equation} \label{constr_ham_n} 
             \boldsymbol\kappa= (2\,{\rm\bf K}^l{}_{l})^{-1}[\,\interior{\rm\bf K}{}_{kl}\,\interior{\rm\bf K}{}^{kl} 
      + 2\,{\rm\bf k}{}^{l}{\rm\bf k}{}_{l}    - \tfrac12\,({\rm\bf K}^l{}_{l})^2 - \hskip-1mm {}^{{}^{(3)}}\hskip-1mm R \,]\,,
\end{equation}
$\hat D_i$ and $\hat R$ denote the covariant derivative operator and scalar curvature associated with
$\hat \gamma_{ij}$, respectively, and $\dot{\hat n}{}_k={\hat n}{}^lD_l{\hat n}{}_k=-{\hat D}_k(\ln{\hat N})$. 
Here (\ref{constr_ham_n}) provides an algebraic solution to the Hamiltonian constraint (\ref{expl_eh})
(for more details see \cite{racz_constraints}).
The four quantities $(\boldsymbol\kappa, {\rm\bf k}{}_{i}, {\rm\bf K}^l{}_{l})$
are subject to the constraints whereas the remaining eight variables
$(\hat N,\hat N^i,\hat \gamma_{ij}, \interior{\rm\bf K}_{ij})$ are
freely-specifiable throughout $\Sigma$. Here $\interior{\rm\bf K}_{ij}$
encodes the two free gravitational degrees of freedom.

Given the free data $(\hat N,\hat N^i,\hat \gamma_{ij}, \interior{\rm\bf K}_{ij})$,
the equations (\ref{constr_mom1})-(\ref{constr_ham_n}) were shown to comprise a first order strongly
hyperbolic system for the vector valued variable $({\rm\bf K}^l{}_{l},{\rm\bf k}{}_{i})$  provided
$\boldsymbol\kappa$ and ${\rm\bf K}^l{}_{l}$ are of opposite sign,
\begin{equation}
       \boldsymbol\kappa{\rm\bf K}^l{}_{l}  = -C^2 \, , \quad C \ne 0.
     \label{eq:sign}
\end{equation}
It was also verified in \cite{racz_constraints} that, given the values of $({\rm\bf k}{}_{i}, {\rm\bf K}^l{}_{l})$
on some ``initial'' surface  $\mycal{S}_0$ satisfying (\ref{eq:sign}), solutions to the nonlinear system 
(\ref{constr_mom2})-(\ref{constr_ham_n}) exist (at least locally) in a neighborhood
of $\mycal{S}_0$, and that the fields $(h_{ij},K_{ij})$ built up from these solutions satisfy
the full constraint system (\ref{expl_eh})-(\ref{expl_em}). 

%%%%%%%%%%%%%%%%%%%%%%%%%%%%%%

\section{The free and constrained Schwarzschild data}

\label{sec:kerr-schild}

The successful application of this new approach to the constraint problem depends
upon a judicious choice of  gauge, determined by the lapse of the initial Cauchy hypersurface $\Sigma$,
and a judicious choice of foliation $\mycal{S}_\rho$. We begin by considering data in Kerr-Schild form,
in which the space-time metric has the form
\begin{equation}%\label{SymS2}
         g_{ab}=\eta_{ab}+2 H \ell_a \ell_b\,,  \quad  g^{ab}=\eta^{ab}-2 H \ell^a \ell^b \, ,
         \label{eq:ksm}
\end{equation}
where $H$ is a smooth function (except at singularities) on $\mathbb{R}^4$ and
$\ell_a$ is null with respect to both $g_{ab}$ and an implicit background
Minkowski metric $\eta_{ab}$. In inertial coordinates $(t,x^i)$ adapted to $\eta_{ab}$,
\begin{equation}
         g_{ab}dx^a dx^b= (-1+2H{\ell_t}^2)dt^2 + 4H \ell_t\ell_i dt dx^i  + (\delta_{ij}+ 2 H \ell_i \ell_j)dx^i dx^j\,, 
\end{equation}
% THE ABOVE EQUATION WAS CORRECTED
where  $\ell^a=g^{ab}\ell_b = \eta^{ab}\ell_b$ and 
$g^{ab}\ell_a\ell_b = \eta^{ab}\ell_a\ell_b=-(\ell_t)^2+\ell^i\ell_i=0$.
The Kerr-Schild metrics also satisfy the background geodesic condition 
\begin{equation}
           \eta^{bc} \ell_c \partial_b  \ell_a=0
           \label{eq:ksg}
\end{equation}
and wave equation 
\begin{equation}
       \eta^{ab}\partial_a \partial_b H =0 .
        \label{eq:ksw}
\end{equation}

We can relate the Kerr-Schild metric to the $3+1$ decomposition of the space-time metric
\begin{equation}
      g_{ab}=h_{ab}- n_a n_b , 
\end{equation}
where $n^a$ is the future directed unit normal to the $t=const$ hypersurfaces.
Choose a time evolution field $t^a$ satisfying $t^a\partial_a t=1$.
Then  $n^a$ has the decomposition 
\begin{equation}%\label{SymS2}
         n^a={N}^{-1} (t^a-N^a)\, , 
\end{equation}
where $N$ and $N^a$ denote the spacetime lapse and shift, determined by 
\begin{equation}%\label{SymS2}
N= -\,(t^e n_e)\, , n_a=-N \partial_a t  \hskip0.5cm {\rm and} \hskip0.5cm N^a={h^{a}}_{e}\,t^e\,,
\end{equation}
respectively.

In the Kerr-Schild spacetime coordinates $(t,x^i)$,  the metric has components
\begin{eqnarray}\label{split_ginv4x4_s}
&&g_{\alpha\beta}=\left(
\begin{array}{cc}
-{N^2}+{N_iN^i} & {N_i} \\
{N_j} & h_{ij} 
\end{array} 
\right)\,.
\end{eqnarray} 
It follows that
%\begin{equation}%\label{SymS2}
%h_{ij}=\delta_{ij}+2 H\ell_i \ell_j\,, \quad h^{ij} =\delta^{ij} -\frac{2 H}{1+2H}  \ell^i \ell^j\, ,
%\end{equation}
%%
%???CORRECTION???
\begin{equation}%\label{SymS2}
 h_{ij}=\delta_{ij}+2 H\ell_i \ell_j\,, \quad h^{ij} =\delta^{ij} -\frac{2 H\ell^i \ell^j}
 {1+2H\ell_t^2}  \, , 
\end{equation}

\begin{equation}%\label{SymS2}
    N=\frac{1}{\sqrt{1+2 H\ell_t^2}}\,, 
\end{equation}
\begin{equation}\label{eq:shift}
N_i=2 H\ell_t \ell_i\,, \quad N^i=2 H N^2\ell_t \ell^i \, .
\end{equation}

\medskip

A direct calculation of the extrinsic curvature
\begin{equation}
K_{ij}=\tfrac12\,\mycal{L}_{n} {h}_{ij}=(2N)^{-1}[\,\partial_{t} {h}_{ij}-(D_iN_j+D_jN_i)]
\end{equation}
gives
\begin{align}
N^{-1} K_{ij}= {}&- \ell_t \left[ \partial_i  (H\ell_j)+\partial_j (H\ell_i \right) ] 
    +N^{-2} \partial_t (H\ell_i \ell_j)    \nonumber \\ 
  {}& +2 H \ell^t \ell^k \partial_k(H\ell_i \ell_j) -H(\ell_i\partial_j \ell_t+\ell_j \partial_i \ell_t)  .
  \label{eq:kerrxc}
\end{align}

For a Kerr spacetime
\begin{equation}
H=\frac{rM}{r^2+a^2\cos^2\theta}\,,
\end{equation}
where the Boyer-Lindquist radial coordinate $r$ is related to  the Cartesian inertial
spatial coordinates $x^i =(x_1,x_2,x_3)$ according to
\begin{equation}
r^2=\frac12\left[\,(\rho^2-a^2)+\sqrt{(\rho^2-a^2)^2+4 a^2 x_3^2}  \, \right]
\end{equation}
with 
\begin{equation}
     \rho^2=x_1^2+x_2^2+x_3^2\, 
     \label{eq:minkrho}
\end{equation}
and
\begin{equation}
       \ell_a=\left(1, \frac{rx_1+ax_2}{r^2+a^2},
       \frac{rx_2-ax_1}{r^2+a^2},\frac{x_3}{r}  \right)\,.
\end{equation}

As $H$ and $\ell_a$ are $t$-independent and $\ell_t=1$,
the extrinsic curvature (\ref{eq:kerrxc}) simplifies to
\begin{equation}
       K_{ij}= - \ell_t N\left[ \partial_i  (H\ell_j)+\partial_j (H\ell_i)
    + 2 H \ell_i \ell_j\ell^k\partial_k H\,\right]  \nonumber \,.
\label{eq:K}
\end{equation}

Formally, for the purpose of applying the approach in Sec.~\ref{sec:setup}
to a generic inspiral and merger, it would
be sufficient to show that the required sign condition (\ref{eq:sign})
holds for a boosted
Kerr black hole. Here we restrict our investigation to the Schwarzschild
case, where the choice of foliation $\mycal{S}_\rho$ is guided by
spherical symmetry and the algebraic simplicity allows a clear
exposition of the approach.

For a Schwarzschild black hole, the spin parameter $a=0$ and
the Kerr-Schild form of the metric simplifies to
\begin{equation}
       H=\frac{M}{r}\,, \quad  \ell_i=\frac{x_i}{r}= \partial_i r\, , \quad r^2 =\delta^{ij} x_1 x_j \, ,
\end{equation}
with lapse 
\begin{equation}
     N=(1+ 2H)^{-1/2}
\end{equation}
and 3-metric
\begin{equation}
       h_{ij}= \delta_{ij} + 2 H \ell_i\ell_j\, .
\end{equation}
(Here $-\ell^a$ is a future directed ingoing null vector, which corresponds
to the convention for ingoing Eddington-Finklestein coordinates.)
Thus
\begin{equation}
     \partial_i H =-\frac{M}{r^3}\,x_i \,, \quad \partial_j (H \ell_i) =\frac{M}{r^4}\left[ r^2 \delta_{ij}- 2 x_ix_j\right]\,
\end{equation}
and (\ref{eq:K}) reduces to
\begin{equation}
    K_{ij}=  -\frac{2 M}{r^2\,\sqrt{1+2H}} \left( \delta_{ij}
   - \left[\,2 +H\,\right] \ell_i\ell_j \,\right) \,.
\end{equation}

We choose the foliation $\mycal{S}_\rho$ by setting $\rho=r$, with
$\rho^i=\ell^i$, corresponding to the ``spatial'' lapse and shift
\begin{equation}\label{Nhat}
       \hat N= \sqrt{1+2H} \, , \quad \hat N^i= 0 \, ,
\end{equation}
unit normal
\begin{equation}
        \hat n_i=\sqrt{1+2H}\,  \ell_i\,, \quad \hat n^i= h^{ij}\hat n_j=\frac{1}{\sqrt{1+2 H  }}\, \ell^i\,,
\end{equation}
and intrinsic 2-metric 
\begin{equation}
      \hat \gamma_{ij} =  h_{ij}-\hat  n_i \hat n_j = \delta_{ij}- \ell_i\ell_j\,, 
          \quad  \hat \gamma^{ij}= \delta^{ij} -\ell^i\ell^j\, .
\end{equation}

A straightforward calculation gives the extrinsic curvature components of $\Sigma$,
\begin{align}
\boldsymbol\kappa = {}&  \hat n^k\hat  n^l\,K_{kl} =
\frac{2 M\,\left( 1+H \right)}{r^2\,\left(1+2H\right)^{3/2}} \, ,  \label{bfkappa}\\
{\rm\bf k}{}_{i} = {}&  {\hat \gamma}^{k}{}_{i}\hat  n^l\, K_{kl} = 0 \, ,\\
{\rm\bf K}_{ij} = {}&  {\hat \gamma}^{k}{}_{i} {\hat \gamma}^{l}{}_{j}\,K_{kl} =
 -\frac{2 M}{r^2\,\sqrt{1+2H}}\, \hat \gamma_{ij} \, ,
\end{align}
\begin{equation}\label{trbfK}
{\rm\bf K}^l{}_{l} = \hat \gamma^{kl} {\rm\bf K}_{kl}  = -\frac{4 M}{r^2\,\sqrt{1+2H}} \, ,
\end{equation}
and 
\begin{equation}
\interior{\rm\bf K}_{ij}={\rm\bf K}_{ij}-\tfrac12\,\hat \gamma_{ij} \,{\rm\bf K}^l{}_{l} = 0\,.
\end{equation}
Note that $\boldsymbol\kappa$ and ${\rm\bf K}^l{}_{l}$ are globally non-vanishing and 
have opposite sign, in agreement with the condition (\ref{eq:sign}) for
strong hyperbolicity.

From (\ref{hatextcurv}) along with
\begin{align}
    \mycal{L}_{\hat n} {\hat \gamma}_{ij} = {}& \hat n^k \partial_k {\hat \gamma}_{ij} 
    + {\hat \gamma}_{kj} (\partial_i\hat n^k)+ {\hat \gamma}_{ik} (\partial_j\hat n^k) \\ 
    = {}& \frac{1}{\sqrt{1+2 H  }}\, \frac{x^k}{r} \partial_k \left[-\frac{x_ix_j}{r^2}\right] 
    +  \left(\delta_{kj}- \frac{x_kx_j}{r^2}\right) \partial_i \left[\frac{1}{\sqrt{1+2 H  }}\frac{x^k}{r}\right] 
         \nonumber \\  {}&
    +  \left(\delta_{ik}- \frac{x_ix_k}{r^2}\right) \partial_j \left[\frac{1}{\sqrt{1+2 H  }}\frac{x^k}{r}\right] 
    = \frac{2}{r\,\sqrt{1+\frac{2M}{r}}}\,\left(\delta_{kj}- \frac{x_kx_j}{r^2}\right) \,, \nonumber
\end{align}
the extrinsic curvature of the $\rho=r=const$ foliated surfaces is given by 
\begin{equation}
     \hat K_{ij}=  \frac{1}{r\,\sqrt{1+2H}}\,{\hat \gamma}_{ij}\,,
\end{equation}
so it follows that
\begin{equation}\label{trKhat}
  {\hat K}{}^l{}_{l} = \hat \gamma^{kl} {\hat K}_{kl}  = \frac{2}{r\,\sqrt{1+2H}} 
\end{equation}
and
\begin{equation}
      \interior{\hat K}_{ij}={\hat K}_{ij}-\tfrac12\,\hat \gamma_{ij} \,{\hat K}{}^l{}_{l} = 0 \, .
\end{equation}

\section{Nonlinear perturbations of a Schwarzschild black hole}
\label{sec:pert}
Here we investigate nonlinear perturbations of the Kerr-Schild initial data for
a Schwarzschild black hole. In doing so, we simplify the discussion by assigning
Schwarzschild values to the freely specifiable
variables $(\hat N,\hat N^i,\hat \gamma_{ij},\interior{\rm\bf K}_{ij})$. 
As a result, the initial $3$-metric $h_{ij}$ retains its Schwarzschild value and,
in particular, $\interior{\rm\bf K}_{ij}=0$ and $\hat N$ and ${}^{{}^{(3)}}\hskip-1mm R$
have no angular dependence.
For a more general perturbation, $(\hat N,\hat N^i,\hat \gamma_{ij},\interior{\rm\bf K}_{ij})$
would enter as explicit terms in the resulting system for
$(\boldsymbol\kappa, {\rm\bf K}^l{}_{l},{\rm\bf k}{}_{i},)$. 

In this setting, (\ref{constr_mom2})-(\ref{constr_mom1}) reduce to 
\begin{align}
\mycal{L}_{\hat n}({\rm\bf K}^l{}_{l}) - \hat D^l {\rm\bf k}_{l} - [\,\boldsymbol\kappa-\tfrac12\, ({\rm\bf K}^l{}_{l})\,]\,
({\hat K^{l}}{}_{l})  = {}& 0 \,,  
     \label{eq:constr_mom2}  \\
\mycal{L}_{\hat n} {\rm\bf k}{}_{i}  + ({\rm\bf K}^l{}_{l})^{-1}[\,\boldsymbol\kappa\,
     \hat D_i ({\rm\bf K}^l{}_{l}) -2\, {\rm\bf k}{}^{l}\hat D_i{\rm\bf k}{}_{l}\,] 
+ ({\hat K^{l}}{}_{l})\,{\rm\bf k}{}_{i}  = {}& 0 ,\label{eq:constr_mom1} 
\end{align}
where $\boldsymbol\kappa$, determined by (\ref{constr_ham_n}), reduces to
\begin{equation} \label{eq:constr_kappa} 
             \boldsymbol\kappa= (2\,{\rm\bf K}^l{}_{l})^{-1}[ 2\,{\rm\bf k}{}^{l}{\rm\bf k}{}_{l}    
             - \tfrac12\,({\rm\bf K}^l{}_{l})^2 - \hskip-1mm {}^{{}^{(3)}}\hskip-1mm R \,]\,.
\end{equation}
It is easy to check that these equations hold for a Schwarzschild solution,
 for which ${}^{{}^{(3)}}\hskip-1mm R = \frac{8 M^2}{r^4 (1+ 2 H)^2}$,
 ${\hat n}{}^i\partial_i=\frac{1}{\sqrt{1+2H}} \partial_r$, ${\rm\bf k}{}_{i}=0$
 and neither ${\rm\bf K}^l{}_{l}$ nor $\boldsymbol\kappa$ have angular
 dependence.
 
In spherical coordinates $x^i =(r,x^A)$, $x^A=(\theta,\phi)$,
\begin{equation}\label{eq:spdc}
     \hat\gamma_{ij}dx^i dx^j=r^2 q_{AB} dx^A dx^B
\end{equation}
where $q_{AB}$ is the unit sphere metric. 
Then (\ref{eq:constr_mom2})-(\ref{eq:constr_mom1}) become
\begin{align}
 \frac{1}{\sqrt{1+2H}} \partial_r {\rm\bf K}^l{}_{l} - \hat D^B {\rm\bf k}_{B} 
    - [\,\boldsymbol\kappa-\tfrac12\, ({\rm\bf K}^l{}_{l})\,]\, ({\hat K^{l}}{}_{l})  = {}& 0 \, ,
      \label{eq:smom2}  \\
    \frac{1}{\sqrt{1+2H}} \partial_r {\rm\bf k}{}_{A}  + ({\rm\bf K}^l{}_{l})^{-1}[\,\boldsymbol\kappa\,
     \partial_A ({\rm\bf K}^l{}_{l}) -2\, {\rm\bf k}{}^{B}\hat D_A{\rm\bf k}{}_{B}\,] 
   + ({\hat K^{l}}{}_{l})\,{\rm\bf k}{}_{A}  = {}& 0 \label{eq:smom1} .
\end{align}

Now consider nonlinear perturbations of Schwarzschild.
We denote by $\delta V = V-V_S$ the deviation of a variable $V$ from
its Schwarzschild value $V_S$. Then (\ref{eq:smom2})-(\ref{eq:smom1})
take the form
\begin{align}
   \frac{1}{\sqrt{1+2H}} \partial_r \delta {\rm\bf K}^l{}_{l}
     - \frac{q^{BC}}{r^2}\partial_C \delta {\rm\bf k}_{B} 
         = {}&F_1  ,  \label{eq:fsmom2}  \\
    \frac{1}{\sqrt{1+2H}} \partial_r\delta  {\rm\bf k}{}_{A}  + \frac{\boldsymbol\kappa}{{\rm\bf K}^l{}_{l}}
     \partial_A \delta{\rm\bf K}^l{}_{l}
      -\frac {2q^{BD} {\rm\bf k}{}_D}{r^2 {\rm\bf K}^l{}_{l}}\partial_A \delta{\rm\bf k}{}_{B}
   ={}&  F_A  ,\label{eq:fsmom1} 
 \end{align}
where $F_1$ and $F_A$ represent lower differential order terms.
This is a coupled quasilinear system for the vector valued variable
$U_\alpha =(u_1,u_A)=(\delta{\rm\bf K}^l{}_{l},\delta{\rm\bf k}{}_{A})$.
The system  (\ref{eq:fsmom2})-(\ref{eq:fsmom1}) has matrix form
\begin{equation}
        \partial_\tau U_\alpha ={{\cal L}_\alpha}^{\beta C} \partial_C U_\beta + F_\alpha,
        \label{eq:L}
\end{equation}
where $\partial_\tau =(1+2H)^{-1/2} \partial_r$, $F_\alpha =(F_1,F_A)$ and
\begin{eqnarray}
           {{\cal L}_1}^{1 C}  &=&0 , \quad {{\cal L}_1}^{B C} =\frac{1}{r^2}q^{BC} ,\\
           {{\cal L}_A}^{1 C} &=&- \frac{\boldsymbol\kappa}{{\rm\bf K}^l{}_{l}} \delta_A^C ,
           \quad  {{\cal L}_A}^{B C} = \frac {2} {r^2 {\rm\bf K}^l{}_{l}}q^{BD}{\rm\bf k}{}_{D} \delta_A^C .
\end{eqnarray}

The requirement that (\ref{eq:L}) is a strongly hyperbolic system~\cite{kreissl,reula} is that there exists a
positive bilinear form $H_{\beta \gamma}$ such that
${\cal L}(\omega)_{\beta \alpha}=H_{\beta \gamma}{{\cal L}_\alpha}^{\gamma C} \omega_C$ is symmetric
for each choice of $\omega_C$. It is straightforward to check that such a symmetrizer
is given by 
\begin{eqnarray}\label{eq:H}
          H_{11} &=& - \frac{{\rm\bf K}^l{}_{l}}{\boldsymbol\kappa}, \quad H_{1A}=0 ,\\
          H_{A1}&=& \frac{2 {\rm\bf k}{}_{A} }{\boldsymbol\kappa} ,
        \quad  H_{AB}=r^2 q_{AB}.
\end{eqnarray} 
The positivity of the symmetrizer for perturbations of Schwarzschild,
\begin{equation}
        H_{\alpha\beta}v^\alpha v^\beta =  - \frac{{\rm\bf K}^l{}_{l}}{\boldsymbol\kappa} (v^1)^2
        +\frac{2}{\boldsymbol\kappa} {\rm\bf k}{}_{A}v^1 v^A + r^2 q_{AB} v^A v^B >0, \quad v^\alpha\ne0,
\end{equation}
follows from the near Schwarzschild approximations
\begin{equation}
     - \frac{{\rm\bf K}^l{}_{l}}{\boldsymbol\kappa} \approx \frac{2(1+2H)}{1+H}, 
             \quad \frac {{\rm\bf k}{}_A} {\boldsymbol\kappa} \approx 0.
\end{equation}
\medskip
Furthermore, the $\omega_A$ independence of $H_{\alpha\beta}$ implies that
the system is symmetric hyperbolic as well as strongly hyperbolic.

Given near Schwarzschild data for $({\rm\bf K}^l{}_{l},{\rm\bf k}{}_{A})$
on a surface $\mycal{S}_R$ surrounding a Schwarzschild black hole,
strong hyperbolicity is a sufficient condition for the system (\ref{eq:smom2} )-(\ref{eq:smom1})
to produce a unique solution of the constraint problem in some neighborhood
of $\mycal{S}_R$. Furthermore, the problem is well-posed so that the solution
depends continuously on the data.
For linearized perturbations the solution extends
globally to $r=0$. 

\section{Future prospects}
\label{sec:conc}

We have shown that the new treatment of the constraints proposed
in~\cite{racz_constraints}
leads to a well-posed constraint problem for nonlinear perturbations
of a Schwarzschild black hole in Kerr-Schild form. 
As is generally the case for nonlinear
problems, the solution is only guaranteed locally in a neighborhood of
the  outer surface $\mycal{S}_R$ on which the data is prescribed. The issue
of a global solution to the nonlinear problem
is best explored by numerical techniques for integrating
the hyperbolic system inward along the $\rho$-streamlines
emanating from $\mycal{S}_R$.

The well-posedness of this problem extends to perturbations representing
a Kerr black hole with small spin and boost.
The question whether it extends further
to a Kerr black hole with maximal spin and arbitrary boost is more
complicated. Its resolution would depend, among other
things, upon a judicious choice of the foliation $\mycal{S}_\rho$
and the $\rho$-streamlines along which the evolution proceeds.
This is akin to choosing the lapse and shift for
a timelike Cauchy evolution.

The ultimate utility of this new approach rests upon its extension
to multiple black holes. Formally, it can be applied to the multiple black
hole problem using a modification of the
superimposed Kerr-Schild data proposed in~\cite{ksm1,ksm2},
which is based upon the ansatz that the
initial three metric for a binary black hole is given by
\begin{equation}
       h_{ij}= \delta_{ij} + 2 H {}^{[1]}\ell_i {}^{[1]}\ell_j{}^{[1]} 
       +2 H {}^{[2]}\ell_i {}^{[2]}\ell_j{}^{[2]},
       \label{eq:bks3}
\end{equation}
where $ H {}^{[n]}$ and $\ell_i {}^{[n]}$ correspond to the Kerr-Schild data
for individual boosted, spinning black holes.
In~\cite{ksm1,ksm2}, the actual
3-metric data is only conformal to (\ref{eq:bks3}), with the
conformal factor chosen to satisfy
the Hamiltonian constraint.

In our new approach to the constraints, it is possible
to retain the superimposed Kerr-Schild initial data in its strict
4-dimensional form 
\begin{equation}
       g_{ab}= \eta_{ab} + 2 H {}^{[1]}\ell_a {}^{[1]}\ell_b{}^{[1]} 
       +2 H {}^{[2]}\ell_a {}^{[2]}\ell_b{}^{[2]},
       \label{eq:bks4}
\end{equation}
where $\ell_a{}^{[n]}$ are null with respect to the background
Minkowski metric. This determines the initial lapse and shift
as well as the initial 3-metric (\ref{eq:bks3}) for an
evolution along the $\rho$-streamlines. It is is an attractive strategy
because it retains much of the algebraic simplicity of the Kerr-Schild
metric, e.g. $\ell_a{}^{[1]}$ and $\ell_a{}^{[2]}$ satisfy the background
geodesic equation (\ref{eq:ksm}),  $H {}^{[1]}$ and $H {}^{[2]}$ satisfy the
background wave equation (\ref{eq:ksg}) and the metric can be
explicitly inverted, although in a more complicated form
than (\ref{eq:ksw}).

Given the background metric (\ref{eq:bks4}), the
Hamiltonian constraint can be imposed to express the extrinsic curvature
component
$\boldsymbol\kappa$ algebraically in terms of ${\rm\bf K}^l{}_{l}$
and explicitly known terms via (\ref{constr_ham_n}).
The extrinsic curvature components
$\interior{\rm\bf K}_{ij}={\rm\bf K}_{ij}-\tfrac12\,\hat \gamma_{ij}\,{\rm\bf K}^l{}_{l}$,
can be freely prescribed, say, by superposition of their individual Kerr-Schild values.
Given a suitable foliation of the initial hypersurface $\mycal{S}_\rho$ and vector field
$\rho^i$, 
the remaining components of the extrinsic curvature data,
${\rm\bf K}^l{}_{l}$ and ${\rm\bf k}{}_{i}$, could then be determined from
the hyperbolic system (\ref{constr_mom2})-(\ref{constr_mom1}) obtained
from the momentum constraint.
The only data necessary are the values of ${\rm\bf K}^l{}_{l}$ and ${\rm\bf k}{}_{i}$
on a large surface $\mycal{S}_R$ surrounding the system. 
The surface data for ${\rm\bf K}^l{}_{l}$ and ${\rm\bf k}{}_{i}$ 
could be prescribed (again tentatively) by the superposition of their
individual Kerr-Schild values.

A major concern in such a scheme is the effect of caustics, where  the ingoing
$\rho$-streamines focus, or a crossover surface $\mycal{S}_X$ where these streamlines
from opposing points of $\mycal{S}_R$ meet. For a single black hole,
the streamlines can be chosen so that 
the caustics and crossovers are inside the apparent horizon,
where the interior can be excised. However, for
binary black hole data, although the caustics can be arranged to lie
inside the black holes, the crossover surface $\mycal{S}_X$ will in
general span the region between them. In that case, unless $\mycal{S}_X$
can be chosen to be a surface of reflection symmetry, as in the case of data
for an axisymmetric  head-on collision, the inward evolution from  $\mycal{S}_R$ 
may produce a discontinuity on $\mycal{S}_X$, i.e. the data induced on
$\mycal{S}_X$ may not be single-valued. 

Considerable numerical experimentation might be necessary to
deal with this issue. The following strategy,
which puts the flexibility of symmetric hyperbolic systems to use,
is only schematic. 
Unlike the iterative global nature of elliptic solvers,
hyperbolic evolution proceeds locally 
along the $\rho$-streamlines and  can be stopped
freely. This can be utilized to adjust the crossover surface, by numerical experimentation,
so that it minimizes
the discontinuity on  $\mycal{S}_X$
along each pair of intersecting $\rho$-streamlines. Then any discontinuity of the  solution
on  $\mycal{S}_X$ might be
removed by averaging. Since the hyperbolic evolution of the constraint system can also proceed
in the outward  $\rho$-direction,
a smooth solution, using this averaged data on $\mycal{S}_X$ can then be extended
outward to $\mycal{S}_R$ .

The simplicity of such a scheme for binary black hole initial data is
extremely attractive.
Whether it can be successfully implemented is again a matter for
numerical study. If  such studies were indeed successful they would lead to
questions of the utmost physical importance: Does the resulting binary black hole
initial data suppress junk radiation? Does it give better control
over the orbital and spin parameters of a binary system? The sole data needed
on a single large surface in the asymptotic region surrounding the system distinguishes
this approach from other solutions to the constraint problem which rely
on elliptic equations. Whether this feature improves the physical content and control
of the initial data
is again a matter for numerical investigation.

\setcounter{equation}{0}

%%%%%%%%%%%%%%%%%%%% ACKNOWLEDGMENTS %%%%%
\section*{Acknowledgments}

The authors are grateful for the kind hospitality of
the Albert Einstein Institute in Golm,
Germany,  where this work was initiated. IR was supported in part
by the Die Aktion \"Osterreich-Ungarn, Wissenschafts- und Erziehungskooperation grant 90\"ou1.
JW was supported by NSF grant PHY-1201276 to
the University of Pittsburgh. 
  
%%%%%%%%%%%%%%%%%%%% REFERENCES %%%%%%%%
%\section*{References}

\end{document}